\documentclass[preprint]{elsarticle}

\usepackage{graphicx}
\usepackage{amssymb}
\usepackage{amsmath}

\makeatletter
\def\ps@pprintTitle{%
 \let\@oddhead\@empty
 \let\@evenhead\@empty
 \def\@oddfoot{}%
 \let\@evenfoot\@oddfoot}
\makeatother
\journal{Physics Letters B}

\usepackage{color}
\usepackage{tikz}

\newcommand{\ev}[1]{\left\langle #1 \right\rangle}
\newcommand{\obs}{\mathcal{O}}
\newcommand{\rD}[1]{\mathrm{D}[#1]}
\newcommand{\rO}{\mathrm{O}}
\newcommand{\e}{\mathrm{e}}

\newcommand{\mb}{{\overline{m}}}
\newcommand{\nf}{{n_f}}

\newcommand{\set}[1]{\{#1\}}
\newcommand{\Wff}{W_{2\mathrm{f}}}

\newcommand{\fig}[1]{Fig.~\ref{#1}}

\newcommand{\sect}[1]{Section~\ref{#1}}

\begin{document}

\begin{frontmatter}

%% Title, authors and addresses

%% use the tnoteref command within \title for footnotes;
%% use the tnotetext command for the associated footnote;
%% use the fnref command within \author or \address for footnotes;
%% use the fntext command for the associated footnote;
%% use the corref command within \author for corresponding author footnotes;
%% use the cortext command for the associated footnote;
%% use the ead command for the email address,
%% and the form \ead[url] for the home page:
%%
%% \title{Title\tnoteref{label1}}
%% \tnotetext[label1]{}
%% \author{Name\corref{cor1}\fnref{label2}}
%% \ead{email address}
%% \ead[url]{home page}
%% \fntext[label2]{}
%% \cortext[cor1]{}
%% \address{Address\fnref{label3}}
%% \fntext[label3]{}

\title{One flavor mass reweighting in lattice QCD}

%% use optional labels to link authors explicitly to addresses:
%% \author[label1,label2]{<author name>}
%% \address[label1]{<address>}
%% \address[label2]{<address>}

\author[phys]{Jacob Finkenrath}
\author[phys]{Francesco Knechtli}
\author[phys,math]{Bj\"orn Leder}

\address[phys]{Department of Physics, Bergische Universit\"at Wuppertal,
               Gaussstr. 20, D-42119 Wuppertal, Germany}
\address[math]{Department of Mathematics, Bergische Universit\"at Wuppertal,
               Gaussstr. 20, D-42119 Wuppertal, Germany}

\begin{abstract}
One flavor mass reweighting can be used in lattice QCD computations
to fine tune the quark masses to their physical values.
We present a new method that utilizes an
unbiased stochastic estimation of the one flavor determinant.
The stochastic estimation 
is based on the integral representation of the determinant of a
complex matrix, which we prove. In contrast to other methods it can also 
be applied in situations where the determinant has a complex phase. The stochastic
error is controlled by determinant factorizations based on mass interpolation and
Schur decomposition. As an example of an application we demonstrate how the
method can be used to tune the up-down quark mass difference.
\end{abstract}

%\begin{keyword}
%% keywords here, in the form: keyword \sep keyword

%% MSC codes here, in the form: \MSC code \sep code
%% or \MSC[2008] code \sep code (2000 is the default)

%\end{keyword}

\end{frontmatter}

%%
%% Start line numbering here if you want
%%
% \linenumbers

%% main text
\section{Bare parameter reweighting}
\label{sec:reweighting}

Reweighting is an old method introduced in \cite{Ferrenberg:1988yz} to obtain results for a
range of parameters by using a single Monte Carlo simulation at one value of the parameters.
Reweighting can also be used to modify the sampling in a Monte Carlo simulation. In both cases
the result for the desired ensemble is obtained by including a reweighting factor in the 
observables. In lattice quantum chromodynamics (QCD) simulations the reweighting factor is
typically a determinant of a ratio of Dirac operators,
e.g.~\cite{Hasenfratz:2008fg,Luscher:2012av,Aoki:2012st,Aoki:2010dy}.
We close a gap, which to our knowledge exists in the literature, by proving an integral 
representation for the determinant of a complex matrix.
We prove the efficiency of this method by applying it to the case of one flavor reweighting
in the quark mass.

The expectation value of an
observable $\obs$ in lattice QCD is given by the path integral
\begin{equation}\label{eq:obs}
 \ev{\obs}_{a} = \frac{1}{Z_{a}}\int\rD{U}\;P_{a}(U)\; \obs(U)\,,
\end{equation}
where the gauge link configuration $U$ receives the weight 
\begin{equation}\label{eq:weight}
 P_{a}(U) = \e^{-S_g(\beta,U)}\,\prod_{i=1}^{\nf} \det(D(U)+m_i)\,.
\end{equation}
The expectation value $\ev{\cdot}_{a}$ is defined at the set of bare parameters 
$$a=\set{\beta,m_1,m_2,\ldots,m_\nf}\,,$$
where \mbox{$\beta=6/g^2$} is the bare inverse gauge coupling
and $m_1,m_2,\ldots,m_\nf$ are the $\nf$ bare mass parameters of the 
sea quarks. Note that in \eqref{eq:obs} the Grassman valued quark fields have been
integrated out 
yielding the product of determinants in \eqref{eq:weight}.
Therein the ``massless'' Dirac operator $D$ is shifted by
the corresponding mass parameter. The normalization $Z$ is fixed
by demanding $\int\rD{U}P(U)/Z \equiv 1$.  

In numerical computations the path integral in Eq.~\eqref{eq:obs} is evaluated
employing Monte Carlo methods. The expectation value is given as the ensemble mean
\begin{equation}\label{eq:obs-mc}
 \ev{\obs}_{a} = \frac{1}{N}\sum_i^{N} \obs(U_i) + \rO(1/\sqrt{N})\,,
\end{equation}
over an ensemble of gauge field configurations $\set{U_i,\, i=1,\ldots,N}$ that
has been generated according to the probability distribution $P_a(U)$.
Since the generation of such ensembles is a computer time intensive sequential
process it is desirable to reuse a generated ensemble to compute the
expectation value at a different set of parameters. This is the idea of reweighting.

The expectation value at a set of
parameters $b=\set{\beta',m_1',m_2',\ldots,m_\nf'}$ different from $a$,
can be expressed in terms of expectation values at $a$ via
\begin{equation}\label{eq:obs-rew}
 \ev{\obs}_{b} = \frac{\ev{\obs W_{a,b}}_a}{\ev{W_{a,b}}_a}\,,
        \quad W_{a,b} = \frac{P_b}{P_a}\,.
\end{equation}
The ratio $W_{a,b}$ is the so-called \emph{reweighting factor}.
Note that the observable $\obs$ is the same on both sides of 
Eq.~\eqref{eq:obs-rew}. In particular, if it explicitly depends on the bare
parameters they have to be the same on both sides.

In the simplest case the two sets differ only in one parameter and identical terms
trivially cancel in the ratio $W_{a,b}$.
Let us distinguish the \emph{beta shift} reweighting factor, where $a$ and $b$ differ
in the inverse bare coupling
\begin{equation}\label{eq:beta-shift}
  W_{\beta,\beta'} = \e^{-(S_g(\beta',U)-S_g(\beta,U))}\,.
\end{equation}
The \emph{one flavor} reweighting factor, where the two sets differ in
the bare mass parameter for quark $i$, is given by
the determinant of the ratio of the corresponding Dirac operators
\begin{equation}\label{eq:one-flavor}
 W_{m_i,m_i'} = \det\frac{D(U)+m_i'}{D(U)+m_i}\,.
\end{equation}
The other cases, where two or more parameters differ, are products of beta shift and
one flavor reweighting factors.

Beside the one flavor reweighting Eq.~\eqref{eq:one-flavor} the \emph{two flavor}
reweighting is of practical importance.
In the most general case it is the product of two 
one flavor reweighting factors $\Wff=W_{m_i,m_i'}W_{m_j,m_j'}$. We distinguish the
 special cases of reweighting
two degenerate quarks $\Wff\equiv W_{m,m'}^2$, and the 
\emph{isospin} reweighting $\Wff\equiv W_\pm = W_{m,m-\Delta m} W_{m,m+\Delta m}$,
where two degenerate quarks are reweighted in opposite directions.
The one and two flavor reweighting may be combined with a beta shift.

Additional reweighting factors occur
when QED effects or a finite chemical potential are considered. Although we present
numerical results exclusively for reweighting in the mass and the coupling, the
proof of the integral representation of a determinant (Section 
\ref{ssec:integral}) is an important general result.

\newcommand{\etad}{\eta^\dagger}
\newcommand{\psid}{\psi^\dagger}
\newcommand{\F}{\mathcal{F}}
\newcommand{\C}{\mathbb{C}}
\newcommand{\re}{\mathrm{Re}}
\newcommand{\im}{\mathrm{Im}}
\newcommand{\Tr}{\mathrm{Tr}}
\newcommand{\Cp}{\C^+}
\newcommand{\Neta}{N_\eta}
\newcommand{\peta}{{p(\eta)}}
\newcommand{\Dm}{\Delta m}
\newcommand{\dm}{\delta m}
\newcommand{\Dp}{D_m}
\newcommand{\gap}{m_{\mathrm{gap}}}

\section{Stochastic estimation}
\label{sec:stochastic}
Since $D$ is a large sparse matrix the determinant in Eq.~\eqref{eq:one-flavor}
has to be computed using stochastic methods. In this section we define 
an unbiased estimator with controlled convergence. 

\subsection{Integral representation}
\label{ssec:integral}
Let $A$ be a complex matrix with
eigenvalues $\lambda(A)$ and
$\eta$ a complex valued vector. In \ref{ap:integral} we prove
\begin{equation}\label{eq:det-integral}
 \frac{1}{\det A} = \int \rD{\eta}\; \e^{-\etad A \eta}
    \quad \text{if}\quad \lambda(A+A^\dagger) > 0\,. 
\end{equation}
The condition $\lambda(A+A^\dagger)>0$ is the necessary and sufficient 
condition for the absolute convergence of the integral in
Eq.~\eqref{eq:det-integral}. It is equivalent to
the condition that the field of values $\F(A)$ is in the right half plane, 
see \ref{ap:integral}.
It implies the weaker condition $\re(\lambda(A))>0$ (see \ref{ap:integral})
but the two conditions are only equivalent for normal matrices.

If the integral in Eq.~\eqref{eq:det-integral} exists it can be evaluated with
Monte Carlo methods. Let $\set{\eta_k,\, k=1,\ldots,\Neta}$ be
an ensemble of random complex vectors
with probability distribution $p(\eta)$. The determinant can then be written as
\begin{equation}\label{eq:det-integral2}
 \frac{1}{\det A} = \ev{\frac{\e^{-\etad A \eta}}{p(\eta)}}_\peta
                  = \frac{1}{\Neta}\sum_{k=1}^{\Neta} \frac{\e^{-\eta_k^\dagger A \eta_k}}{p(\eta_k)}
                     + \rO(1/\sqrt{\Neta})\,,
\end{equation}
where $\ev{\obs}_\peta = \int \rD{\eta}\; p(\eta)\obs(\eta)$ in accordance with
Eq.~\eqref{eq:obs}. It is convenient to choose a Gaussian distribution
$p(\eta)=\exp({-\etad\eta})$, which we will assume from now on. The ensemble mean in
Eq.~\eqref{eq:det-integral2} is an estimator of the inverse of the determinant and it 
converges if the integral in Eq.~\eqref{eq:det-integral} exists. The variance of this
estimator is given by
\begin{align}
 \sigma_\eta^2 & = \ev{\frac{\e^{-\etad (A+A^\dagger) \eta}}{p(\eta)^2}}_\peta
               - \ev{\frac{\e^{-\etad A \eta}}{p(\eta)}}_\peta
                 \ev{\frac{\e^{-\etad A^\dagger \eta}}{p(\eta)}}_\peta \label{eq:variance}\\
               & = \frac{1}{\det(A+A^\dagger-I)} - \frac{1}{\det(AA^\dagger)}\,.
\end{align}
The first integral in Eq.~\eqref{eq:variance} exists if
\begin{equation}
\label{eq:stoch-var-cond}
 \lambda(A+A^\dagger-I)>0 \Leftrightarrow \lambda(A+A^\dagger)>1 \Rightarrow \re(\lambda(A))>0.5\,.
\end{equation}
The condition $\lambda(A+A^\dagger)>1$ automatically implies the
existence of the second and third integral, cf. Eq.~\eqref{eq:det-integral}. 
Therefore,
if in an implementation of the estimator in Eq.~\eqref{eq:det-integral2} 
the variance \eqref{eq:variance} is
monitored, its convergence assures the convergence of the mean.

Often the matrix $A$ can be written in the form $A=I+\epsilon B$ with
$\epsilon||B||\ll 1$. Then a useful approximation of
the variance \eqref{eq:variance} is obtained by expanding
 in $\epsilon$
\begin{equation}
\label{eq:variance-expand}
   \frac{\sigma_\eta^2}{|\det A|^{-2}} = \det\left( I + \epsilon^2\frac{BB^\dagger}{I+\epsilon(B+B^\dagger)}\right) -1 
     = \epsilon^2 \Tr(BB^\dagger) + \rO(\epsilon^3)\,,
\end{equation}
where we used $AA^\dagger=I+\epsilon(B+B^\dagger)+\epsilon^2 BB^\dagger$
 and $\det(X)=\exp(\Tr\ln(X))$ for positive definite $X$.

\subsection{One flavor reweighting}

The one flavor reweighting factor \eqref{eq:one-flavor} is given by the determinant
of the ratio of two Dirac operators. Dropping the subscripts and explicit dependence on 
the gauge field from now on, and
defining the reweighting distance $\Dm=m-m'$ and $\Dp=D+m$, the reweighting factor
$W_{m,m'}$ can be written as
\begin{equation}\label{eq:one-flavor2}
   W = \frac{1}{\det M}\,, \quad M^{-1} = {\Dp}^{-1}D_{m'}=I - \frac{\Dm}{\Dp}\,.
\end{equation}
Because the eigenvalues of $M+M^\dagger$ converge to two for $\Dm\to0$, 
the integral representation of $W$ in Eq.~\eqref{eq:one-flavor2} exists 
for sufficiently small 
$\Dm$ and non-singular $\Dp$ and $D_{m'}$.
Since $M=I+\epsilon B$ with $\epsilon=\Dm$ and
$B=\Dp^{-1} + \rO(\Dm)$ the variance \eqref{eq:variance} of
stochastically estimating $W$ via \eqref{eq:det-integral2} can be approximated
with Eq.~\eqref{eq:variance-expand}
\begin{equation}
   \frac{\sigma_\eta^2}{|W|^2} = \Dm^2 \Tr((\Dp\Dp^\dagger)^{-1}) + \rO(\Dm^3)\,.
\end{equation}
In contrast to conventional methods
\cite{Aoki:2009ix,Aoki:2012st} the estimator also works in situations where
the determinant is negative or complex (for example, it becomes negative when
a real eigenvalue of $D$ crosses zero between the shifts $m$ and $m'$).

\newcommand{\Dl}{D_{m_l}}

\subsection{Mass interpolation}
\label{ssec:mass-interpolation}
An estimator where the stochastic variance can be reduced (and its convergence controlled)
at fixed $\Dm$ is obtained when
the matrix $M$, and thus the determinant,
is factorized $M = \prod_{l=0}^{N-1} M_l$. 
One possible factorization is given by taking the $N$th root $M_l=M^{1/N}$ \cite{Hasenfratz:2002ym}. Another
possibility that does not involve the evaluation of a matrix function is an interpolation between
$m$ and $m'$ \cite{Hasenfratz:2008fg}. In the context of the Hybrid Monte Carlo algorithm a similar method is known
as ``mass preconditioning'' \cite{Hasenbusch:2001ne}, in the context of the
Partially Stochastic Multi-Step algorithm as ``gauge link interpolation'' \cite{Finkenrath:2012az}.
The idea is to introduce intermediate reweighting factors with a
reweighting ``distance'' that scales with $1/N$. In the case of mass reweighting
we set $\dm=\Dm/N=(m-m')/N$ and
\begin{equation}\label{eq:mass-inter-ratio}
   M_l^{-1} = \frac{D_{m_{l+1}}}{\Dl} = I - \frac{\dm}{\Dl}\,, \quad m_l = m-l\dm\,.
\end{equation}
%Note that $D_0=\Dp$ and $D_{N-1}-\dm=D_{m'}$.
The estimate of the whole
reweighting factor in Eq.~\eqref{eq:one-flavor2} is now given as a product of estimates
\begin{equation}\label{eq:mass-inter}
   W = \prod_{l=0}^{N-1}\; W_l\,, \quad W_l =
    \ev{\frac{\e^{-{\eta^{(l)}}^\dagger M_l \eta^{(l)}}}{p(\eta^{(l)})}}_{p(\eta^{(l)})}\,,
\end{equation}
where for each factor $W_l$ an independent estimator is used.
The variance of each factor can be approximated using Eq.~\eqref{eq:variance-expand}.
If in the Monte Carlo process for each factor $\Neta$ random vectors are used
(cf. Eq.~\eqref{eq:det-integral2}) the \emph{relative stochastic error}
 $\delta_\eta^2 = \sigma_\eta^2/(|W|^2N_\eta)$ of the product is given by
\begin{equation}\label{eq:stochastic-error}
   \delta_\eta^2 = \frac{N}{\Neta}\left[\dm^2 \Tr\left((D_m D_m^\dagger)^{-1}\right) + \rO(\dm^2\Dm)\right]
                           \approx \frac{\Dm^2}{N\Neta}\Tr\left((D_m D_m^\dagger)^{-1}\right)\,,
\end{equation}
where in the second step we assumed the higher order terms to be negligible.
This can always be achieved by choosing $N$ large enough (i.e., $\dm$ small enough)
and then Eq.~\eqref{eq:stochastic-error} states that increasing $N$ or
$\Neta$ has the same effect on decreasing the stochastic error.
Note that the above arguments only hold if the estimate for each factor
is converging. Again, as long as $\Dl$ does not become singular,
convergence can be assured by choosing $N$ large enough, i.e.,
squeezing $\lambda(M_l)$ in a small region around one.

\subsection{Schur decomposition}

\newcommand{\Mee}{M_{\mathrm{ee}}}
\newcommand{\Moo}{M_{\mathrm{oo}}}
\newcommand{\Dee}{D_{\mathrm{ee}}}
\newcommand{\Deo}{D_{\mathrm{eo}}}
\newcommand{\Doo}{D_{\mathrm{oo}}}
\newcommand{\Doe}{D_{\mathrm{oe}}}
\newcommand{\Ds}{\hat{D}}
\newcommand{\smin}{\sigma_{\mathrm{min}}}

Yet another factorization that reduces the stochastic noise can be employed if the lattice
Dirac operator $D$ only involves next-neighbor couplings. In this case the lattice sites
can be labeled \emph{even} and \emph{odd}, depending on whether the sum of their lattice
coordinates is even or odd. Reordering the entries in $D$ such that all even sites come first
the determinant of $D$ becomes a product
\begin{equation}
   \det (D+m) = \det (\Dee+m)\, \det (\Doo+m)\, \det \Ds_m\,.
\end{equation}
of the determinants of the (block) diagonal $\Dee+m$ and $\Doo+m$ and the Schur complement
$\Ds_m=I-(\Dee+m)^{-1}\Deo(\Doo+m)^{-1}\Doe$. The two former ones can be evaluated exactly.
The latter acts non-trivially only on the even lattice sites
and therefore $\dim(\Ds_m) = \dim(D)/2$. The one flavor reweighting factor employing
even-odd factorization and stochastic estimation of the Schur complement ratio with
mass interpolation reads
\begin{equation}\label{eq:mass-inter-eo}
   W = \frac{1}{\det\Mee\,\det\Moo}\;\prod_{l=0}^{N-1}\; \hat{W}_l\,, \quad \hat{W}_l =
    \ev{\frac{\e^{-{\eta^{(l)}}^\dagger \hat{M}_l \eta^{(l)}}}{p(\eta^{(l)})}}_{p(\eta^{(l)})}\,,
\end{equation}
where $\Mee=(\Dee+m)/(\Dee+m')$ and similarly for $\Moo$.
 Noting that the Schur complement
can be expanded $\Ds_{m+\dm} = \Ds_{m} + \dm X_m + \rO(\dm^2)$ with
\begin{equation}
   X_m = - (\Dee+m)^{-1}\left[\Ds_m - I - \Deo(\Doo+m)^{-2}\Doe\right]\,,
\end{equation}
the ratio $\hat{M}_l$ can be cast in a form similar to Eq.~\eqref{eq:mass-inter-ratio}
\begin{equation}
 \hat{M}_l^{-1}=\frac{\Ds_{m_{l+1}}} {\Ds_{m_l}}=I-\dm \Ds^{-1}_{m_l} X_{m_l} + \rO(\dm^2)\,.
\end{equation}
The relative stochastic error of the product $\prod_l\; \hat{W}_l$ is then approximated by
\begin{equation}\label{eq:stochastic-error-eo}
   \delta_\eta^2 \approx \frac{\Dm^2}{N\Neta}\Tr\left(X_m X_m^\dagger(\Ds_m \Ds_m^\dagger)^{-1}\right)\,.
\end{equation}

\subsection{Zero crossings}
\label{ssec:zero_crossings}

%
%%%%%%%%%%%%%%%%%%%%%%%%%%%%%%%%%%%%%%%%%%%%%%%%%%%%%%%%%%%%%%%%%%%%%%%%%%%%%%%
\begin{figure}[t]
%  \begin{center}
   \centering
   \includegraphics*[width=.7\linewidth]{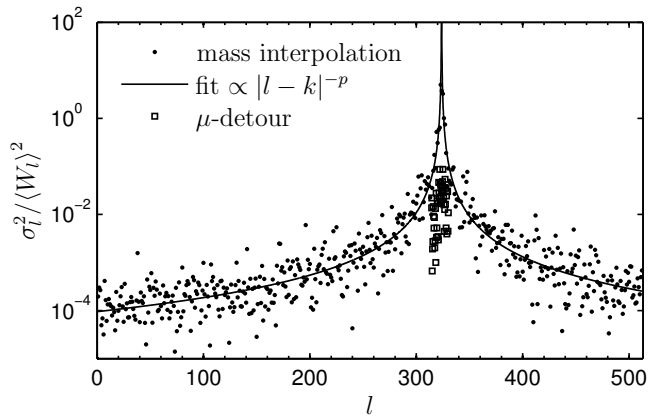}
%  \end{center}
  \caption{\small Example of a zero crossing. The plot shows the variances of the
   factors of mass interpolation along a straight line from $m$ to $m'<m$ 
   in $N=512$ steps (dots), a fit to the data pointing to a divergence at $k\approx 322$ (line)
   and the variances along the regulating $\mu$-detour (squares).
}
  \label{fig:variance_mu}
\end{figure}
%%%%%%%%%%%%%%%%%%%%%%%%%%%%%%%%%%%%%%%%%%%%%%%%%%%%%%%%%%%%%%%%%%%%%%%%%%%%%%%
%
In \fig{fig:variance_mu} we show the variance of the one flavor reweighting factor
for one configuration of the ensemble D5 with mass $m$. Along the mass interpolation 
with $N=512$ steps (dots) to reweight to the mass $m'$ (of D6) a real eigenvalue of $\Dl$
crosses zero, as is indicated by the ``peak'' in the variance. We refer to Section \ref{sec:ensemble}
for the explanations of the ensemble parameters.
For this analysis we used a domain decomposition 
with $6^4$ blocks and the variance is for the stochastic estimation of the reweighting factor
of the Schur complement only (cf. Eq.~\eqref{eq:mass-inter-eo} and comment on
domain decomposition in Section \ref{ssec:mass-summary}). The number of mass interpolation
steps is chosen such that a good resolution of the peak is achieved.
A fit of the variances to $\propto |l-k|^{-p}$ yields a good description of the data (line) with 
$k\approx 322$ and $p\approx 1.8$. An exponent $p=2$ is expected if the
trace in Eq.~\eqref{eq:stochastic-error-eo} is dominated by one small eigenvalue.

The behavior of the variance can be regularized by a detour in the mass-$\mu$ plane
(squares) as explained in \ref{ap:mu-detour}. 
The crucial observation is the well known fact that the singular values of
$D_{m,\mu}=D_m+i\mu\gamma_5$ are bounded from below by $|\mu|$.
The detour starts at $l_s=314$ and ends at $l_e=330$.
The maximal value of $a\mu$ is 0.00003 and the detour is divided into $48$ steps.
The reweighting factor is complex for $\mu\neq0$ and we plot the sum of the variance 
of the real and imaginary parts. As expected by construction the variance is bounded from
above and never larger than $10^{-1}$ along the detour.

Finally we remark that although the single factors along the detour are complex, their
product is real within errors. 
The imaginary parts add up to a phase equal to minus one giving the correct
sign of the one flavor reweighting factor.

\subsection{Summary}
\label{ssec:mass-summary}

Summarizing, the stochastic estimator of the one flavor reweighting factor Eq.~\eqref{eq:one-flavor}
using factorizations based on even-odd and mass interpolation is given by Eq.~\eqref{eq:mass-inter-eo}.
The expectation values of the individual factors $\hat{W}_l$
with respect to $p(\eta^{(l)})$ are obtained via Monte Carlo integration as in Eq.~\eqref{eq:det-integral2}
with a fixed $N_\eta$ for all $l$. Estimates of the variances of the factors are easily obtained
if $N_\eta\geq 6$. If a zero crossing is observed (variance diverges) the mass interpolation is
modified as described in Section \ref{ssec:zero_crossings}. Writing the trace in
Eq.~\eqref{eq:stochastic-error-eo} as $k_\eta V$ (with the lattice volume $V$)
and if higher orders in $\dm$ are negligible, the relative stochastic error is given by
\begin{equation}\label{eq:stochastic-error2}
   \delta_\eta^2 \approx k_\eta\frac{\Dm^2V}{N\Neta}\,.
\end{equation}
 
We close this section by listing some practical remarks:
\begin{itemize}
 \item From the results of \cite{Finkenrath:2012cz} we deduce that higher order terms are negligible for
       $\dm||D_m^{-1}|| \approx \Dm/(\mb N) \lesssim 1/16$, where the smallest eigenvalue of $D_m$ is
       approximated with the renormalized quark mass $\mb$. In Fig.~1 of Ref.~\cite{Finkenrath:2012cz}
       higher order terms are negligible for $N\gtrsim 8$ and $\Dm=\mb/2$, leading to the stated relation.
 \item The number of inversions of the global Dirac operator is given by $N\cdot N_\eta$.
 \item The value of $k_\eta$ with even-odd factorization is roughly a factor two smaller than without.
       Instead of the presented even-odd factorization, a factorization based on a domain decomposition of
       the lattice is also possible. However, the additional gain is small \cite{Finkenrath:2012cz}
       and has to be compared to the additional cost for the needed (exact) block determinants.
 \item In the case of Wilson fermions, $D^\dagger=\gamma_5 D \gamma_5$ and it follows that the determinant
       $\det M^{-1}$ is real. If the determinant is known to be real an improved estimator can be defined by
       replacing $\ev{\cdot}_\peta\to\ev{\re(\cdot)}_\peta$ in Eq.~\eqref{eq:det-integral2}.
In practice this yields an improvement if $\Neta$ is small.
The real part cannot be taken if we replace $D$ by $D+i\mu\gamma_5$ as discussed
in Section \ref{ssec:zero_crossings}.
 \item The vectors $A\eta^{(k)}$ that are computed in Eq.~\eqref{eq:det-integral2} also appear
       in the estimator for $1/\det(AA^\dagger)$. Therefore, if $D^\dagger=\gamma_5 D \gamma_5$, the
       reweighting factors  $W$ and $W^2$ are obtained at once.
       
\end{itemize}

\newcommand{\MeV}{\,\mathrm{MeV}}
\newcommand{\ms}{\mb_{\rm s}}
\newcommand{\mup}{m_{\rm u}}
\newcommand{\md}{m_{\rm d}}
\newcommand{\ks}{\kappa_{\rm s}}
\newcommand{\ku}{\kappa_{\rm u}}
\newcommand{\kd}{\kappa_{\rm d}}

\section{Ensemble fluctuations}
\label{sec:ensemble}

We present results for reweighting of an ensemble generated with the Wilson
gauge action and two flavors of O($a$) improved Wilson fermions at parameters
$\beta=5.3$, $c_{\rm sw}=1.90952$, $\kappa=0.13625$
on a $48\times24^3$ lattice. The lattice spacing is $a=0.0658(10)$ \cite{Fritzsch:2012wq}.
We use the hopping parameter $\kappa$ which is related
to the bare quark mass $m$ by $\kappa=1/(2am+8)$.
The parameters and boundary conditions are the same as the ones of the D5
 simulation of \cite{DelDebbio:2007pz}.
The pseudoscalar mass is $m_{\rm PS} \simeq 440 \MeV$ and the renormalized quark mass is
$\mb({\rm D5})\approx \ms/3$ \cite{Fritzsch:2012wq}, where $\ms$ is the physical
 strange quark mass. We generated a statistics of
2012 configurations, separated by $2.0$ molecular dynamics units, using mass preconditioned
HMC \cite{Marinkovic:2010eg}.
The target renormalized quark mass $\mb'$ is a factor two smaller, i.e., $\mb'\approx \ms/6$.
It corresponds to the parameter $\kappa'=0.136350$
or $m'_{\rm PS} \simeq 310 \MeV$, as in the D6 simulation of \cite{Luscher:2012av}.

After analyzing the ensemble fluctuations of the one flavor reweighting factor we show
how correlated reweighting factors can be combined in order to reduce the ensemble fluctuations.
In particular, we present the combination of one and two (degenerate) flavor reweighting
with a beta shift and combinations of different one flavor reweighting factors.

\newcommand{\kf}{k_{\rm 1f}}
\newcommand{\sigf}{\sigma_{\rm 1f}}

\subsection{One flavor reweighting}
\label{ssec:one-flavor}
%
%%%%%%%%%%%%%%%%%%%%%%%%%%%%%%%%%%%%%%%%%%%%%%%%%%%%%%%%%%%%%%%%%%%%%%%%%%%%%%%
\begin{figure}[t]
   \centering
   \includegraphics*[width=.7\linewidth]{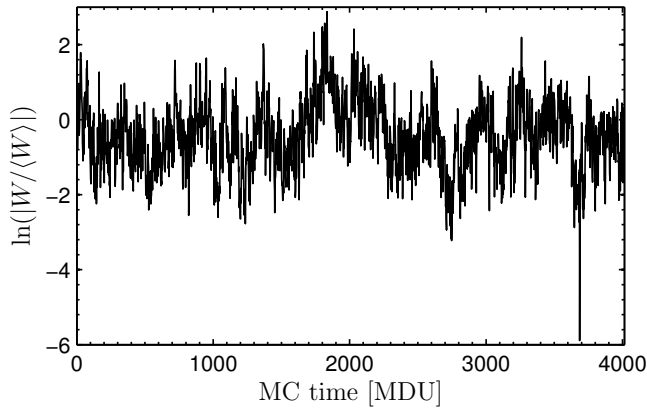}
  \caption{\small History of $\ln(\left|W/\ev{W}\right|)$ for reweighting
  from D5 to D6 using mass interpolation and $\mu$-detour (where necessary).}
  \label{fig:hist_wMu}
\end{figure}
%%%%%%%%%%%%%%%%%%%%%%%%%%%%%%%%%%%%%%%%%%%%%%%%%%%%%%%%%%%%%%%%%%%%%%%%%%%%%%%
%

The one flavor mass reweighting factor Eq.~\eqref{eq:one-flavor2} can be written as
\begin{equation}\label{eq:one-flavor3}
W = \exp\left[\Tr\,\ln\,\left(I-\frac{\Delta m}{D_m}\right)\right]\,.
\end{equation}
Expanding for small $\Delta m$, 
the relative ensemble fluctuations of the reweighting factor can be shown to be
\begin{equation}\label{eq:one-flavor-fluct}
\frac{\sigf^2}{\ev{W}^2} =\frac{\ev{W^2}}{\ev{W}^2}-1 = \Dm^2\left[\ev{(\Tr(D_m^{-1}))^2}
-\ev{\Tr(D_m^{-1})}^2\right] + {\rm O}(\Dm^3) \,.
\end{equation}
In \fig{fig:hist_wMu} we plot the history of the one flavor reweighting
factor, evaluated on each configuration with the techniques explained in 
\sect{sec:stochastic}. The configuration with a very small reweighting factor corresponds to 
an exceptional configuration \cite{Bardeen:1997gv,DeGrand:1998mn} (without zero crossing), 
on which the pseudoscalar propagator has (almost) a pole. This pole is regulated
by the reweighting factor.

Numerically we find that the ensemble fluctuations scale proportional to the lattice volume $V$
for the full Dirac operator (there are indications for a weaker volume 
dependence for the Schur complement only),
cf.~Fig.~2 of Ref.~\cite{Finkenrath:2012cz}. Thus the ensemble fluctuations can be described by
\begin{equation}\label{eq:one-flavor-fluct2}
 \frac{\sigf^2}{\ev{W}^2} \approx \kf \Dm^2 V\,,
\end{equation}
and the leading dependence on the lattice volume $V$ and the reweighting distance $\Dm$
cancels in the ratio of relative stochastic error Eq.~\eqref{eq:stochastic-error2} and 
ensemble fluctuations Eq.~\eqref{eq:one-flavor-fluct2}. 
Eventually we will require that this ratio is at the level of 10\%. In order
to achieve this we have to determine how many inversions $N\Neta$ are necessary. 
Assuming
\begin{equation}\label{eq:ratio-stochastic-ensemble}
 \frac{\ev{\delta_\eta^2}}{\sigf^2/\ev{W}^2} \equiv \frac{\ev{k_\eta}}{\kf N\Neta}\,,
\end{equation}
the ratio $\ev{k_\eta}/\kf$ can be determined from the measured relative stochastic
 error and the ensemble
fluctuations for different $V$, $\Dm$ and renormalized quark masses $\mb$. We find it to be
independent of the lattice volume and to mildly change between $2$ and $3$ if $\Dm$ and $\mb$
are varied in the ranges $0\leq \Dm \leq m/2$ and $\ms/6\leq \mb \leq 4/3\cdot\ms$.\footnote{%
For $\mb\neq \mb(\text{D5})\approx \ms/3$ this
means a partially quenched determination of the reweighting factor.}
The value of $\kf$, and thus the fluctuations, can be reduced by exploiting the correlation
between one flavor and beta shift reweighting factor, which we explain next.

Finally we remark that in the computation of the statistical error of re\-weighted
observables the correlation between the observable and the reweighting factor in
Eq.~\eqref{eq:obs-rew} has to be taken into account and this requires some care.
For example, we use \cite{Wolff:2003sm} where we incorporated the analytic derivatives
of the quotient in Eq.~\eqref{eq:obs-rew}.

\subsection{Beta shift}
%
%%%%%%%%%%%%%%%%%%%%%%%%%%%%%%%%%%%%%%%%%%%%%%%%%%%%%%%%%%%%%%%%%%%%%%%%%%%%%%%
\begin{figure}[t]
  \centering
   \includegraphics*[width=.7\linewidth]{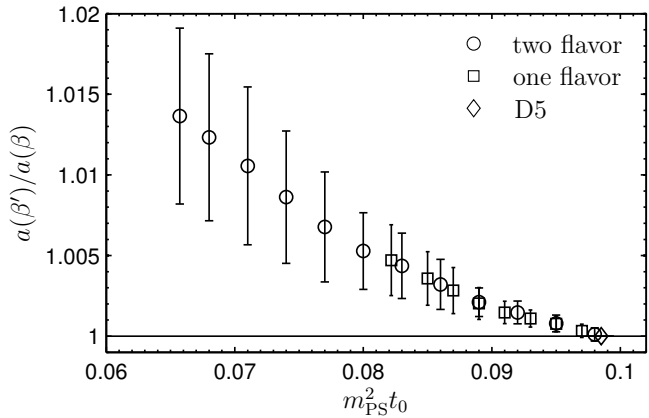}
  \caption{\small Relative change of the lattice spacing with beta shift, one
  and two (degenerate) flavor reweighting.}
  \label{fig:a_shift}
\end{figure}
%%%%%%%%%%%%%%%%%%%%%%%%%%%%%%%%%%%%%%%%%%%%%%%%%%%%%%%%%%%%%%%%%%%%%%%%%%%%%%%
%

We introduce a beta shift, cf. Eq.~\eqref{eq:beta-shift}, by minimizing the
variance of
\begin{equation}\label{eq:beta-shift2}
 \ln(W_{\beta,\beta'}W)=-\Delta\beta S_g(U)
 +\Tr\,\ln\,\left(I-\frac{\Delta m}{D_m}\right) \,,
\end{equation}
as a function of $\Delta\beta=\beta'-\beta$.
For the two (degenerate) flavor case, one can see from Eq.~\eqref{eq:beta-shift2}, by making
an expansion for small mass shifts $\Dm$, that $\Delta\beta$ is to first approximation
a function of the sum of the masses of the two flavors. 
Furthermore, for $N_{\rm f}$ mass degenerate 
flavors $\Delta\beta$ is proportional to $N_{\rm f}$. 
Numerically we find $\Delta\beta/N_{\rm f}\simeq-3\times10^{-4}$
for $\Dm=m/2$.\footnote{The change in the clover coefficient
 $c_{\rm sw}$ \cite{Jansen:1998mx} is of the same 
order of magnitude and can be therefore neglected.}
The fluctuations $\sigma_{\Delta\beta}^2/\ev{W_{\Delta\beta}}^2$ of the reweighting factor 
$W_{\Delta\beta}=W_{\beta,\beta'}W^{N_{\rm f}}$ are reduced by roughly
a factor two with respect to $W^{N_{\rm f}}$ alone, i.e., we find 
$k_{\Delta\beta}/\kf\approx 0.4$.

Our result means that the ensemble fluctuations are minimized if
the lattice spacing is increased when decreasing the quark mass(es). The effect is
shown in \fig{fig:a_shift}. In order to quantify it we use the reference scale $t_0$
introduced in \cite{Luscher:2010iy}. The change of the lattice spacing is given by
$a(\beta')/a(\beta)=\sqrt{t_0(\beta)/t_0(\beta')}$ and is plotted 
as a function of
the non-singlet pseudoscalar mass $m_{\rm PS}^2t_0$ in units of $t_0$, for $N_{\rm f}=1$ and
$N_{\rm f}=2$. We observe that the lattice spacing $a$ changes by 1\%-2\% at most
and therefore lies within the accuracy of $a$ quoted in \cite{Fritzsch:2012wq}.

%The beta shift reduces the range of quark masses, which can be 
%reached by reweighting. 
Since the lattice spacing grows with the beta shift, the reweighting distance
$\Dm$ in physical units is less than it is without beta shift.
We also observed
that the beta shift keeps the value of the plaquette constant within errors.

\newcommand{\Dmp}{D_{m_+}}
\newcommand{\sff}{\sigma_{2\mathrm{f}}^2}

\subsection{Two flavor reweighting}

Correlations can also be exploited when two quark flavors $r$ and $s$ are reweighted simultaneously.
Without loss of generality let the masses $m_r\leq m_s$ be 
reweighted by $-\gamma\Dm$ and $\Dm$, respectively. Defining $m_\pm=(m_s\pm m_r)/2$
the reweighting factor $\Wff = W_{m_r,m_r-\gamma\Dm} W_{m_s,m_s+\Dm}$ can be
written as
\begin{equation}
\label{eq:2f_reweighting}
 \Wff^{(\gamma)}  
     =  \det\left[I-\Dm\frac{(\gamma-1)\Dmp+(\gamma+1)m_-+\gamma\Dm}{\Dmp^2-m_-^2}\right]\,. 
\end{equation}
If the two quarks are degenerate $m_r=m_s=m$ and thus $m_+=m$ and $m_-=0$, three
special values of $\gamma$ occur. For $\gamma=0$ and $\gamma=-1$ one flavor
 reweighting $\Wff^{(0)}=W_{m,m'}$ and two degenerate flavor reweigthing
 $\Wff^{(-1)}=W_{m,m'}^2$
 are recovered, respectively (with $m'=m+\Dm$). The case $\gamma=1$
equals the isospin reweighting $\Wff^{(1)}=W_\pm$ below,
 cf.~Eq.~\eqref{eq:isospin_reweighting}.
The ensemble fluctuations in the three cases at the same $\Dm$ fulfill
$\sigma_{2\mathrm{f},-1}^2 > \sigma_{2\mathrm{f},0}^2 > \sigma_{2\mathrm{f},1}^2$.
The first relation is obvious and the second is a consequence of the suppression
with the fourth power of $\Dm$ in Eq.~\eqref{eq:isospin-fluct}.

Imposing $\gamma=1$ in the non-degenerate case
means keeping the sum of the bare quark masses constant, but is not the optimal
choice in terms of fluctuations. Instead the value of $\gamma$ might
be fixed by minimizing the ensemble fluctuations of $\Wff$. 
From the preceding discussion
we expect the minimum for $0\leq \gamma^* \leq 1$, depending on $m_-$:
if $m_-=0$ isospin reweighting and thus $\gamma=1$ is optimal, whereas if $m_s\to\infty$
(and thus $m_-\to\infty$)
one flavor reweighting of $m_s$ and thus $\gamma=0$ is optimal.
%A calculation similar to Section \ref{ssec:one-flavor} yields
We write Eq.~\eqref{eq:2f_reweighting} as
\begin{equation}
 \Wff^{(\gamma)} = \exp(w)\,,\quad w=-\Dm[(\gamma-1)\Tr\,b + (\gamma+1)\Tr\,c]+
\rO(\Dm^2) \,,
\end{equation}
with $b=\Dmp/(\Dmp^2-m_-^2)$, $c=m_-/(\Dmp^2-m_-^2)$.
Minimization of the variance
${\rm var}(w)=\ev{w^2}-\ev{w}^2$ with respect to $\gamma$ yields
\begin{equation}
 \gamma^* \approx \frac{{\rm var}(\Tr\,b)-{\rm var}(\Tr\,c)}{{\rm var}(\Tr\,b + \Tr\,c)} \,.
\end{equation}
Approximating, for small $m_-$,
${\rm var}(\Tr\,b)\approx {\rm var}(\Tr\,\Dmp^{-1}) \approx \kf V$ and
${\rm var}(\Tr\,c)\approx m_-^2{\rm var}(\Tr\,\Dmp^{-2}) \approx m_-^2k_\pm V$
and neglecting the covariance ${\rm cov}(\Tr\,b,\Tr\,c)$ we obtain
\begin{equation}
\label{eq:optimal-gamma}
   \gamma^* \approx 1- 2m_-^2\frac{k_\pm}{\kf} + \rO(\Dm,m_-^3)\,. 
\end{equation}
We have determined $\gamma^*$ directly by minimizing the fluctuations of
Eq.~\eqref{eq:2f_reweighting} for $m_-\approx\ms/2$ and $m_+\approx 5/6\,\ms$. 
Inserting in Eq.~\eqref{eq:optimal-gamma} $m_-= 50 \MeV$ \footnote{
Differences of quark masses renormalize with a factor $Z_{\rm m}$ which is
$\approx1.5$ in our case \cite{Fritzsch:2012wq} and we neglect it here.}
and the value of $k_\pm/\kf$,
independently determined below for $m_+\approx\ms/3$, results in $\gamma^*\approx 0.86$.
In the direct determination we find the optimal value $\gamma^*=0.82(1)$ in good
agreement with the prediction.

We now return to the special case $m_-=0$, $\gamma=1$.
It means reweighting two degenerate quarks with mass $m=m_r=m_s$
by keeping the sum of the bare masses $m_r+m_s$ constant. The resulting
\emph{isospin} reweighting factor is
\begin{equation}\label{eq:isospin_reweighting}
 W_\pm = W_{m_r,m_r-\Dm} W_{m_s,m_s+\Dm} = \det\left[1-\left(\frac{\Dm}{\Dp}\right)^2\right]
 \,.
\end{equation}
Following the same calculation which led to Eq.~\eqref{eq:one-flavor-fluct},
 we obtain in this case for the fluctuations
\begin{equation}\label{eq:isospin-fluct}
\frac{\sigma_\pm^2}{\ev{W_\pm}^2} = \Dm^4\left[\ev{(\Tr(\Dp^{-2}))^2}
-\ev{\Tr(\Dp^{-2})}^2\right] + {\rm O}(\Dm^6) \,,
\end{equation}
which shows a suppression with the fourth power of $\Dm$, as compared to the second power in
Eq.~\eqref{eq:one-flavor-fluct}.
%, and a leading term given by the trace of $\Dp^{-2}$.
If we use mass interpolation with \footnote{Even-odd preconditioning of $\Dp$ is not
advantageous here since it would lead to terms of order $\dm$.}
\begin{equation}
 M_l^{-1} = \frac{D_{m_{-(l+1)}}}{D_{m_{-l}}}\frac{D_{m_{l+1}}}{\Dl} = I - \frac{(1+2l)\dm^2}{D_m^2-l^2\dm^2}\,,
\end{equation}
the same is true for the relative stochastic error, such that
the ratio of the both is again independent of $\Dm$ and $V$.
Numerically we find $\ev{k_\eta}/k_\pm\approx 2.5-5$ for reweighting distances up to
 $m_s'-m_r'=m/4$. Comparing at the same total reweighting distance
isospin reweighting to the one flavor reweighting of Section \ref{ssec:one-flavor}
we find
$k_\pm/\kf\,(m_s'-m_r')^2= 2.5(6)\times 10^{2}\,a^2(m_s'-m_r')^2$.
Which means that at the physical up-down quark
mass difference, cf.~Section \ref{sec:tuning}, the ensemble fluctuations are a factor $10^{-4}$
smaller for the isospin reweighting.

\newcommand{\mKo}{m_{{\rm K}^0}}
\newcommand{\mKpm}{m_{{\rm K}^\pm}}
\newcommand{\FKo}{f_{{\rm K}^0}}
\newcommand{\FKpm}{f_{{\rm K}^\pm}}

\section{Tuning of the up-down quark mass difference}

In lattice QCD computations the $\nf$ bare mass parameters are usually related to a
minimal number $n_R$ of dimensionless ratios $R_j$, $j=1,\ldots,n_R$ of observables.
The minimal
number is given as the number of independent mass parameter: For example, if $\nf=2$ and
$m_1\equiv m_2$ (a doublet of degenerate quarks) then $n_R=1$. The bare masses are then
tuned such that these ratios take their physical value $R_{j,{\rm phys}}$.
In practice this amounts to the
generation of several ensembles at different values of the bare masses and extrapolation
or interpolation to the physical point. Even if the tuning of $R_j$ is independent of
the tuning of $R_i$ for $i\neq j$ the numerical cost grows rapidly with $n_R$.
Reweighting can help to reduce the number of necessary ensembles.

We present two tuning strategies for the parameters $\ku$, $\kd$ and $\ks$ using
the kaons $K^0$ and $K^\pm$. More precisely, we complement the strategy 
of Ref.~\cite{Fritzsch:2012wq} to fix $\ks$
at degenerate light quarks (therein called strategy 1)
by two different strategies to fix the up-down quark mass difference.
The dimensionless ratios to be used are
\begin{equation}\label{eq:R1-3}
   R_1 = \frac{\mKo^2}{\FKo^2}\,, \quad R_2 = \frac{\mKo^2-\mKpm^2}{\mKo^2}\,,
    \quad R_3=\frac{m_{\pi^0}^2}{\FKo^2}\,.
\end{equation}
with physical values
$R_{1}^{\rm phys}=(494.2)^2/(155)^2$, $R_{2}^{\rm phys}=(497.2^2-491.2^2)/(497.2^2)$ and
$R_{3}^{\rm phys}=(134.8)^2/(155)^2$,  where QED effects have been removed as 
discussed in \cite{Colangelo:2010et}. In the limit $\mup=\md$ all the kaons are the same. 
A mass difference $\md-\mup\neq0$ breaks isospin and splits the kaon masses $\mKo\neq\mKpm$.

\subsection{One flavor reweighting}
\label{sec:tuning}

%
%%%%%%%%%%%%%%%%%%%%%%%%%%%%%%%%%%%%%%%%%%%%%%%%%%%%%%%%%%%%%%%%%%%%%%%%%%%%%%%
\begin{figure}[t]
  \centering
   \includegraphics*[angle=0,width=.7\linewidth]{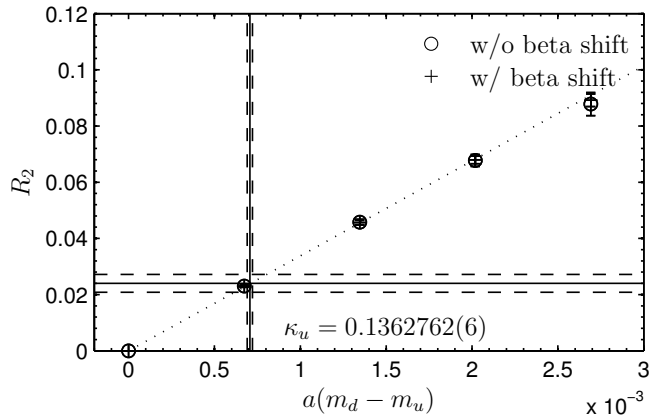}
  \caption{\small Tuning the mass difference $a(\md-\mup)$ to fulfill $R_2\equiv R_{2,{\rm phys}}$,
            cf. Eq.~\eqref{eq:R1-3}.
}
  \label{fig:tuning_m_u}
\end{figure}
%%%%%%%%%%%%%%%%%%%%%%%%%%%%%%%%%%%%%%%%%%%%%%%%%%%%%%%%%%%%%%%%%%%%%%%%%%%%%%%
%
In the first strategy we use one flavor reweighting to lower the mass of the up quark $\mup$.
The outline of the strategy is:
\begin{enumerate}
 \item For $\ku=\kd$, tune $\ks$ such that $R_1\equiv R_{1,{\rm phys}}$.
 \item Keeping $\kd$ and $\ks$ fixed, tune (increase) $\ku$ such that $R_2\equiv R_{2,{\rm phys}}$.
 \item Repeat steps 1. and 2. for different values of $\kd$ and extrapolate to the
 physical quark mass values, e.g., to the point where $R_3\equiv R_{3,{\rm phys}}$.
\end{enumerate}

Step 1 has been performed in \cite{Fritzsch:2012wq} with a quenched strange quark
and step 3 is beyond the scope of this paper.
\fig{fig:tuning_m_u} shows the result of step 2 performed on
the ensemble D5, cf. Section \ref{sec:ensemble}. The parameter $\ks=0.135777$ is taken from 
Table 1 (ensemble E5) of \cite{Fritzsch:2012wq} and the mass $\mup$ of the up quark is
reweighted down to $1/2$ of the down quark mass. We also have computed the reweighting
factors at three intermediate points, allowing for
an interpolation. The ratio $R_2$ of Eq.~\eqref{eq:R1-3} is shown in \fig{fig:tuning_m_u}
and a linear interpolation (corresponding to leading order of chiral perturbation theory)
 yields $\ku=0.1362762(6)$ or $a(\md-\mup)=0.00071(2)$. 
This is approximately $1/4$ of the maximal reweighting distance ($\mb/2$) in \fig{fig:tuning_m_u} or
$4\MeV$ and amounts to $6$\% lower average renormalized quark mass.

For this procedure to work, the ratio $R_1$ in Eq.~\eqref{eq:R1-3} should not significantly change
when tuning $\ku$, i.e., the conditions $R_1\equiv R_{1,{\rm phys}}$ and
 $R_2\equiv R_{2,{\rm phys}}$ should be approximately independent. Indeed, we find the changes in $R_1$
to be much smaller than its statistical error.
Finally we observe that the numerical result for $a(\md-\mup)$ is
practically the same when the beta shift is included in the tuning, cf. \fig{fig:tuning_m_u}.

\subsection{Isospin reweighting}
\label{sec:second_tuning}

We modify step 2 of the tuning described in \sect{sec:tuning} to change simultaneously $\mup$
and $\md$ by keeping their sum constant. In this case we have to replace $R_1$ in Eq.~\eqref{eq:R1-3}
by
\begin{equation}\label{eq:R1_modified}
 R_1' = 0.5 \left(\frac{\mKo^2}{\FKo^2}+\frac{\mKpm^2}{\FKpm^2}\right) \,,
\end{equation}
which is in leading order of chiral perturbation theory a function of the light quark mass sum $\mup+\md$.
$R_1'$ is equal to $R_1$ for $\mup=\md$ and thus step 1 is unchanged.
Using this alternative tuning procedure in step 2 we obtain on the ensemble D5
$\ku=0.1362633(3)$ and $\kd=0.1362367(3)$ or $a(\md-\mup)=0.00072(1)$.

The normalized reweighting factor $W_\pm/\ev{W_\pm}$ close to the tuned mass difference 
is very close to one: deviations are $\lesssim 10^{-4}$. This is mainly due to the 
suppression of ensemble fluctuations with the fourth power of the mass difference in
isospin reweighting (Eq.~\eqref{eq:isospin-fluct}). Consequently, repeating the tuning
with $W\equiv 1$ gives practically the same result and we conclude that at this average
light quark mass isospin breaking can be treated in the quenched approximation \cite{deDivitiis:2013xla}.

However, this will most probably change as soon as smaller light quark masses are considered.
That is because in the tuning step 2 the small eigenvalues of $D$ will be shifted close to zero
or even cross zero (cf.~Section \ref{ssec:zero_crossings}). At this point the inclusion of the reweighting
factor will become indispensable to compensate poles in the propagators.

\section{Conclusion}
\label{sec:conclusion}

The interest in reweighting in the lattice QCD community has been growing in
recent years, but a comprehensive analysis was missing.
The main achievement of this paper is to close this gap by:
\begin{itemize}
 \item Giving a proof for the integral representation of the inverse complex 
       determinant of a complex matrix, valid if the field of values of the 
       matrix is in the right half plane.
 \item Defining an unbiased stochastic estimator based on mass interpolation
       and optional $\mu$-detour
       that has a controlled error and works also in situations where the determinant
       has a complex phase.
 \item Expansions of the relative stochastic error and the ensemble fluctuations
       that are shown to correctly describe the data and lead to optimized re\-weighting
       strategies. 
\end{itemize}

Because of Eq.~\eqref{eq:ratio-stochastic-ensemble} the numerical cost for keeping
the relative stochastic error at a level of 10\% of the ensemble fluctuations is
independent of the lattice volume and the ensemble quark mass: ca.~$50$ inversions
of the global Dirac operator. The reweighting distance is restricted by the
growth of the ensemble fluctuations in Eq.~\eqref{eq:one-flavor-fluct2},
 also known as ``overlap problem''.

As an example of an application we presented two approaches to the tuning of
isospin breaking. 
The first reweights the up quark to a smaller mass, thus lowering the
light quark mass average $(\mup+\md)/2$. This introduces
ensemble fluctuations  which can be tamed by a simultaneous beta shift.
In the second approach the tuning is done at fixed $\mup+\md$. This
leads to a suppression of the ensemble fluctuations at the price of a larger final
average quark mass.

Further possible applications are the tuning of the strange quark mass in simulations
with dynamical up, down and strange quark, stabilization of the Hybrid Monte Carlo
algorithm, treatment of determinants with phases and the inclusion of QED effects.

\smallskip

{\bf Acknowledgments.}  We thank Martin L{\"u}scher for pointing out
a mistake in Eq.~\eqref{eq:det-integral}.
The program package which we developed is based on
the freely available software package DD-HMC
\cite{Luscher:2005rx,Luscher:2007es,Luscher:ddhmc}
and we thank the ALPHA collaboration for
providing the MP-HMC code and the CLS consortium for providing the code for
measuring the scale $t_0/a^2$.
This work was partially funded by Deutsche
Forsch\-ungs\-gemeinschaft (DFG) in form of Transregional 
Collaborative Research Centre 55 (SFB/TRR55).
The numerical calculations were carried out on the
Cheops supercomputer at the RRZK computing centre of the University of Cologne
and on the cluster Stromboli at the University of Wuppertal and we thank both
Universities.

%% The Appendices part is started with the command \appendix;
%% appendix sections are then done as normal sections
\appendix

\section{}
\label{ap:integral}
Let $A\in \C^{n\times n}$ with eigenvalues $\lambda(A)$ and $\eta\in\C^{n}$.
Let $f(\eta)$ be a function $f: \C^{n} \to \C$. We define the integral of
$f$ over $\eta$ through
\begin{equation}
   \int \rD{\eta}\;f(\eta)\quad \text{with}\quad
    \rD{\eta} = \prod_i^n\,\frac{d\re(\eta_i)\,d\im(\eta_i)}{\pi}\,.
\end{equation}
With this definition $\int \rD{\eta}\;\e^{-\etad\eta}=1$.
In order to prove Eq.~\eqref{eq:det-integral}, we first note that the
integral over $f(\eta)=\exp(-\etad A \eta)$ is defined if and only if 
the integral over the absolute value 
$|f(\eta)|=\exp(-\re(\etad A \eta))$ is defined. The condition of
absolute convergence of the integral is therefore equivalent to
the condition that the field of values $\F(A)$ defined in Eq.~\eqref{eq:fov}
is in the right half plane or that $\lambda(A+A^\dagger)>0$ (see below).
Now consider the Schur decomposition of $A$, i.e.,
$A=QUQ^{-1}$ with unitary matix $Q$ ($Q^{-1}=Q^\dagger$) and upper triangular matrix
$U$. The diagonal entries of $U$ are the eigenvalues $\lambda(A)$.
Therefore $U$ can be written
as $U=D+K$ with diagonal matrix $D=\mathrm{diag}(\lambda(A))$ and 
strictly upper triangular matrix
$K$. Since $Q$ is unitary the substitution $\eta\to Q \eta$ in the integral leads to 
\begin{equation}
 \int \rD{\eta}\; \e^{-\etad A \eta}
    = \int \rD{\eta}\; \e^{-\etad (D+K) \eta}\,.
\end{equation}
Now we introduce the two vectors
\begin{equation}
 r=\frac{1}{2} [\eta + (I+D^{-1}K^T)\eta^*]\quad \text{and}\quad
    s=-\frac{i}{2} [\eta - (I+D^{-1}K^T)\eta^*]\,.
\end{equation}
A few lines of algebra show that $\etad (D+K) \eta = r^T D r + s^T D s$. In terms of
the real part $x=\re(\eta)$ and imaginary part $y=\im(\eta)$ the vectors $r$ and $s$ read
\begin{equation}\label{eq:changeofvariables}
 r=x + \frac{1}{2} D^{-1}K^T(x-iy)\quad \text{and}\quad
    s=y + \frac{i}{2} D^{-1}K^T(x-iy)\,.
\end{equation}
Because $D^{-1}K^T$ is strictly lower triangular, the Jacobian matrix $M$ of the change
 of variables
\begin{equation}
 \begin{pmatrix} x\\ y \end{pmatrix} 
 \to \begin{pmatrix} r\\ s \end{pmatrix}=M\begin{pmatrix} x\\ y \end{pmatrix}
 \,,\quad
 M = \begin{pmatrix}
   I+\frac{1}{2} D^{-1}K^T & -\frac{i}{2} D^{-1}K^T\\
   \frac{i}{2} D^{-1}K^T & I+\frac{1}{2} D^{-1}K^T
 \end{pmatrix}
\end{equation}
has determinant one. Thus we obtain
\begin{equation}\label{eq:proof-det-integral}
 \int \rD{\eta}\; \e^{-\etad A \eta}
    = \int \left(\prod_i^n\,\frac{dr_i\,ds_i}{\pi}\right)\; \e^{-r^T D r - s^T D s} 
    = \prod_i^n \frac{1}{\sqrt{\lambda_i} \sqrt{\lambda_i}} = \frac{1}{\det A}\,.
\end{equation}
The transformations in Eq.~\eqref{eq:changeofvariables} change 
the real variables $x=\re(\eta)$ and $y=\im(\eta)$ into
the complex variables $r$ and $s$. The domain of the integrals over the
components of $r$ and $s$ in Eq.~\eqref{eq:proof-det-integral} can be chosen 
to be along the real axis only if the integral is absolute convergent.
The latter property implies that $\re(\lambda(A))>0$ and therefore
the existence of the Gaussian integrals in the last step of
Eq.~\eqref{eq:proof-det-integral}.
We thus complete the proof of Eq.~\eqref{eq:det-integral}.

Note that if $A$ is Hermitian the condition $\lambda(A+A^\dagger)>0$
is equivalent
to $A$ being positive definite and thus the Cholesky decomposition $A=LL^\dagger$
exists. In this case the determinant of $A$ can be shown to appear as the Jacobian
determinant of $\eta\to (L^\dagger)^{-1}\eta$.

We close this Appendix by proving inequality Eq.~\eqref{eq:stoch-var-cond}. The field
of values of $A$ is a subset of the complex plane and defined by
\begin{equation}\label{eq:fov}
   \F(A) =\set{\etad A \eta:\, \etad\eta=1}\,.
\end{equation}
In the proof we will use that (see for example \cite{Parker:fieldofvalues})
\begin{align}
   \F(A) &\supset\lambda(A) \label{eq:fov-prop1}\\
   \F(A) &=[\min(\lambda(A)),\max(\lambda(A))]\quad \Leftrightarrow \quad A=A^\dagger\,.\label{eq:fov-prop2}
\end{align}
Let $d>0$, then
\begin{align*}
 \lambda(A+A^\dagger) > d & \stackrel{\eqref{eq:fov-prop2}}{\Leftrightarrow} \F(A+A^\dagger) > d 
     \stackrel{\eqref{eq:fov}}{\Leftrightarrow} \etad (A+A^\dagger) \eta > d\,,\quad \forall \etad\eta=1 \\
   & \stackrel{\phantom{(A.9)}}{\Leftrightarrow} 2\re(\etad A \eta) > d\,,\quad \forall \etad\eta=1 \\
   & \stackrel{\eqref{eq:fov}}{\Leftrightarrow} 2\re(\F(A)) > d
     \stackrel{\eqref{eq:fov-prop1}}{\Rightarrow} 2\re(\lambda(A))>d 
\end{align*}
proves Eq.~\eqref{eq:stoch-var-cond} for $d=1$. \hfill $\Box$

\section{}
\label{ap:mu-detour}

The stochastic variance of the factors $W_l$ in Section \ref{ssec:mass-interpolation}
diverges if $\Dl$ becomes singular. In order to derive an approximation
of the smallest singular value of $D_{m_l,\mu}=\Dl+i\mu\gamma_5$, let us assume
without loss of generality that $D_{m_{k}}$ is singular for a value $k$ in the
 range $0<k\leq N-1$.
The smallest singular value $\smin$ of $D_{m_l,\mu}=\Dl+i\mu\gamma_5$ is the square root of
the smallest eigenvalue of $D_{m_l,\mu}^\dagger D_{m_l,\mu} = \mu^2 + \Dl^\dagger\Dl$.
Using Eq.~\eqref{eq:min_ev} of \ref{ap:spectral_flow} we arrive at
\begin{equation}
   \smin \approx \sqrt{\mu^2 + (l-k)^2\dm^2\chi_0^2}\,.
\end{equation}
In practice, the so called \emph{chirality} (cf.~Eq.~\eqref{eq:chirality}) $\chi_0$
of the zero mode of $\gamma_5 D_{m_k}$
is very close to plus or minus one \cite{Edwards:1998gk,Simma:1997us} and we assume
here $\chi_0^2=1$.

The stochastic variance is bounded from above if $\smin\geq r$ for some $r>0$. This
is achieved by a detour in the mass-$\mu$ plane. Instead of
going from $m$ to $m'$ on a straight line, a half circle
parametrized by $\mu_j^2 + (\widetilde{m}_j-m_k)^2=r^2$, $j=0,\ldots, N_c-1$ is
 inserted around $m_k$.
The starting point is then given for $\mu_0=0$ and $\widetilde{m}_0=m_{l_s}$
 with $l_s=k-r/|\dm|$
and can be fixed by restricting the increase of the variance, say by a factor four compared 
to the variance far away from $k$.
Equidistant steps
along the half circle are obtained if the reweighting distances
  $\dm_c=\widetilde{m}_{j+1}-\widetilde{m}_j$ and
$\delta \mu_c=\mu_{j+1}-\mu_j$ fulfill $\dm_c^2+\delta\mu_c^2 = \dm^2$ in each step.
We note that for a constant $\chi_0^2<1$ the circle becomes an ellipse with 
the staring point given by $l_s=k-r/|\dm\chi_0|$.

Finally we remark that the detour in the mass-$\mu$ plane can be performed in the same way
with the even-odd (or any domain decomposition) Schur complement, i.e., by considering 
 $\Ds_{m_l,\mu}=I-(\Dee+m_l+i\mu\gamma_5)^{-1}\Deo(\Doo+m_l+i\mu\gamma_5)^{-1}\Doe$.
If $D_{m_l,\mu}$ is assured to be non-singular, the same holds for $\Ds_{m_l,\mu}$.

\section{}
\label{ap:spectral_flow}

Let $D$ be a lattice Dirac operator with the property $\gamma_5 D \gamma_5 = D^\dagger$.
Then $Q(m)=\gamma_5(D+m)$ is Hermitian with real eigenvalues $\lambda_i(m)$ and corresponding
orthonormal eigenvectors $v_i(m)$. The first derivative of the spectral flow is given
by $d\lambda_i/dm=\chi_i$ with the so called \emph{chirality} of the eigenvector
\begin{equation}\label{eq:chirality}
 \chi_i\equiv v_i^\dagger \gamma_5 v_i\,. 
\end{equation}
A Taylor expansion of the spectral flow around $m_0$
at first order is then given by $\lambda_i(m) \approx \lambda(m_0) + (m-m_0)\chi_i(m_0)$
\cite{Edwards:1998gk,Simma:1997us}. 

Let us assume without loss of generality that $Q(m)$ has a zero eigenvalue for $m=m_0$,
i.e., $\min_i|\lambda_i(m_0)|=0$.
Assuming the order of the eigenvalues not to change in the range of interest,
the smallest eigenvalue of $(D+m)^\dagger(D+m)=Q^2$ is
\begin{equation}\label{eq:min_ev}
 \lambda_{\mathrm{min}}=(\min_i|\lambda_i(m)|)^2\approx (m-m_0)^2\chi_0^2\,,
\end{equation}
where $\chi_0$ is the chirality of the zero mode of $Q(m_0)$.

%% References
%%
%% Following citation commands can be used in the body text:
%% Usage of \cite is as follows:
%%   \cite{key}         ==>>  [#]
%%   \cite[chap. 2]{key} ==>> [#, chap. 2]
%%

%% References with bibTeX database:

\bibliographystyle{elsarticle-num}
\bibliography{one-flavor}

%% Authors are advised to submit their bibtex database files. They are
%% requested to list a bibtex style file in the manuscript if they do
%% not want to use elsarticle-num.bst.

%% References without bibTeX database:

% \begin{thebibliography}{00}

%% \bibitem must have the following form:
%%   \bibitem{key}...
%%

% \bibitem{}

% \end{thebibliography}

\end{document}